\documentclass[aps,prl,preprint,groupedaddress]{revtex4-2}
\usepackage{graphicx}
\usepackage{chemformula}
\usepackage{siunitx}
\DeclareSIUnit{\molar}{M}

\begin{document}


\title{Ultra Electron Density Sensitivity for Surface Plasmons}


\author{Wei Liu}
\thanks{These authors contributed equally to this work.}
\affiliation{%
Center for Gravitational Wave Experiment, Institute of Mechanics, Chinese Academy of Science, 15 Bei-si-huan West Road, Beijing, 100190, China.
}%
\author{Meng Li}
\thanks{These authors contributed equally to this work.}
\affiliation{%
Key Laboratory of Analytical Chemistry for Life Science of Shaanxi Province, School of Chemistry and Chemical Engineering, Shaanxi Normal University, Xi'an, 710062, China.
}%
\author{Yu Niu}
 \email{niuyu@imech.ac.cn}
 \affiliation{%
Center for Gravitational Wave Experiment, Institute of Mechanics, Chinese Academy of Science, 15 Bei-si-huan West Road, Beijing, 100190, China.
}%
\author{Ziren Luo}
 \email{luoziren@imech.ac.cn}
\affiliation{%
Center for Gravitational Wave Experiment, Institute of Mechanics, Chinese Academy of Science, 15 Bei-si-huan West Road, Beijing, 100190, China.
}%


\date{\today}

\begin{abstract}
We investigate surface plasmons from a solid-state standpoint and highlight their ultra electron density sensitivity.  When a surface plasmon is excited on a planar gold film by an evanescent wave from 625 nm light, only a minute fraction of the surface electron density, approximately one thousandth, participates in the process. By introducing a noise-depressed surface potential modulation, we reduce the electron density to the order of \qty{10}{\per\square\um}, enabling electron sensitivity on the order of 0.1 e. As a practical application, we develop a surface plasmon resonance imaging method capable of detecting single anions in solution at a concentration of \qty{1}{\atto\molar}.  
\end{abstract}


\maketitle

The label-free single molecule imaging technique has come into the spotlight for it can not only yield the insights about the dynamic molecule interactions that are fundamentally inaccessible to fluorescence-based methodologies but also enjoy the multiplexing capability \cite{Non-fluorescentReview}. Several techniques based on the light-scattering phenomenon cross the barrier of the large mismatch between the size of the molecule and the diffraction limit of visible light successfully. Interferometric scattering microscopy (iSCAT), a recently developed technique \cite{iSCATPRL2004,iSCATOE2006} which collects the scattered photons across a molecule and reduces the background scattering, is able to provide the molecular weight of the molecule in solution \cite{iSCATScience}. By measuring the inelastic light-scattering process, surface-enhanced Raman scattering (SERS) \cite{SERSPRL1997,SERSScience1997} and tip-enhanced Raman spectroscopy (TERS) \cite{stockle2000TERS} can tell the Raman "fingerprint" of the target molecule since a high degree of structural information about the molecule can be extracted from the SERS/TERS vibrational spectrum. Although these scattering-based methods have achieved a remarkable success, the scattering efficiency is restricted by the effective scattering cross section of the molecule. The optical resolution of iSCAT is claimed around \qty{10}{\nm} at room temperature in solution while that of TERS is about \qtyrange{3}{15}{\nm} in vacuum. Thus, it is still a challenge to observing single molecules beyond the restriction of the size under the condition when the target molecules are functionally active.

Besides the molecular size, other properties come into the picture of the single-molecule investigation among which reactions occurring at the single-molecule level is at the core. For chemical reactions, molecules transfer or share electrons at their specific electronic states: the highest occupied molecular orbital (HOMO) and the lowest unoccupied molecular orbital (LUMO). Holding electrons at the highest energy, the HOMO could transfer electrons to other molecules while the LUMO remains empty and capable of receiving electrons. Therefore, a possible strategy to bypass the scattering cross section restriction is to measure the electron density variations induced from the electronic states of the molecules at the sensing surface.

Under total internal reflection condition, the electric field of an evanescent wave can excite the collective oscillation of surface electrons, which is called surface plasmon resonance (SPR). Conventionally, a biosensor based on SPR is described by a Four-layer model governed by Snell's law: glass substrate/ plasmonic metal film/ molecular film/ buffer medium. Billions of molecules adsorbing at the sensing surface changes the thickness of the molecular film which can be measured by a SPR biosensor. However, when the concentration of the target molecules comes down to single-molecule level, less than \qty{1}{\pico\molar} \cite{wu2018challenges}. The model fails because it is hardly to take the sparsely adsorbate molecules as a film. As a result, the detection limit of a SPR biosensor has been considered far from the requirement of the single molecule detection. A typical refractive index detection limit for a phase-sensitive SPR scheme is reported as the order of $10^{-8}$ to date \cite{kabashin2009phase} while the refractive index variation from \qty{1}{\pico\molar} ions is estimated as $10^{-14}$ in solution \cite{leyendekkers1977refractive}. To exceed the limit, plasmon-enhanced nano-materials \cite{ament2012single,zijlstra2012optical}, meta-surface \cite{zeng2015graphene}, and whispering-gallery-mode \cite{vollmer2008whispering,dantham2013label,baaske2014single} are introduced to increase the sensitivity of the biosensor. Although these micro/nano-structure based solutions reveal exciting sensitivity, their applications have some fundamental restrictions. For one thing, the signal amplitude of these methods crucially depends on the location where the molecule binds with the structure while these binding events are outside the control. For another, the large-scale and reproducible production of the complicated sensing unit imposes a technical limitation. Here, we investigate surface plasmons from a solid-state perspective and demonstrate the ultra electron sensitivity of a planar SPR film to measure surface electron density variations. As an application, we have developed an an energy level aligned SPR imaging (ELA-SPRi) system to detect the adsorption and desorption of single anions in solution. 

\begin{figure}[ht]
\centering
\includegraphics[width=0.9\textwidth]{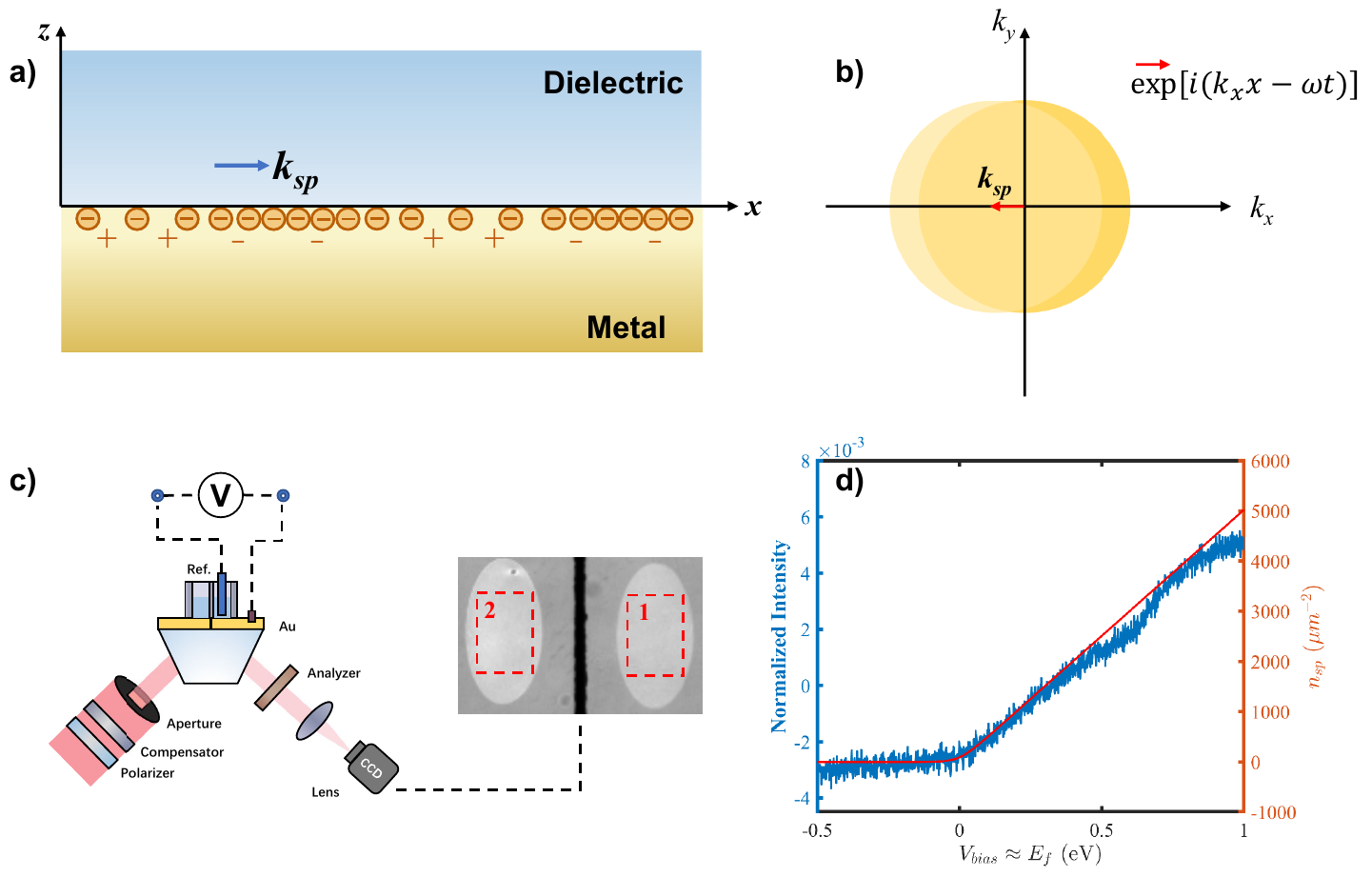}
\caption{(a) The collective oscillation of a surface plasmon with the wave-vector $k_{sp}$ at a metal-dielectric interface. (b)  In momentum space,  a unit evanescent wave, $\exp{i(k_x x-\omega t)}$, can excite the collective oscillation of a surface plasmon when  $k_x=k_{sp}$. Only electrons near the Fermi surface participate in the oscillation. (c) Schematics of the phase sensitive SPR imaging system with an external signal generator which can apply a bias potential to the sensing surface. (d) The measured SPR signal (blue line) and theoretical calculation of $n_{sp}$ (red line) under a linear potential from \qtyrange{-0.5}{1}{\volt}.}\label{fig:1}
\end{figure}

The sensitivity of SPR to the surface charge has been reported since 1970s \cite{abeles1975investigation, kotz1977electron,foley2008surface,abayzeed2017sensitive} and used to detect charged particles at the sensing surface \cite{shan2010measuring}.  This sensitivity is believed from the agreement that the dielectric function of the metal is determined by the electron density of the metal and an applied bias potential on the surface can modulate the electron density \cite{liu2016potential,abayzeed2017sensitive}. However, this understanding overlooks the plasmonic side of the story. From solid-state perspective, the collective oscillation of surface electrons in Fig.~\ref{fig:1}(a) with the wave-vector of the surface plasmon, $k_{sp}$, occurs near the Fermi surface of the metal. In other words, the electron density involved in surface plasmon, $n_{sp}$, is only a fraction of the background surface electron density, $n_s$, in momentum space in Fig.~\ref{fig:1}(b), and can be given by
\begin{equation} \label{Eq:Nsp}
    n_{sp} = \frac{k_{sp}}{k_F}n_{s}
\end{equation}
where $k_F$ is the Fermi wave-vector for gold, around \qty{1.21d4}{\per\um}. The resonance condition, the match between the wave-vectors of the evanescent wave, $k_x$, and that of the surface plasmon, $k_{sp}$, gives the estimation of $k_{sp}/k_F$ about $10^{-3}$, since  $k_{sp} = k_x\approx 2\pi/\lambda\approx$ \qty{10}{\per\um} for the visible light with the wavelength $\lambda =$ \qty{0.625}{\um}.

Eq.~\ref{Eq:Nsp} suggests the SPR signal be an indicator to the surface electron density. At the noble-metal surface, the surface electron density follows a two-dimensional electron gas (2DEG) model by \cite{ziman1962electrons}:
\begin{equation} \label{Eq:N2d}
    n_s = n_{2D} = \frac{m}{\pi\hbar^2}k_BT\ln\biggl( 1+\exp{\frac{E_f}{k_BT}}\biggr)
\end{equation}
where  $m$ is the electron mass, $\hbar$ Planck's constant, $k_B$ Boltzmann constant, $T$ the temperature and $E_f$ the Fermi level of the metal. 

According to Eqs.~\ref{Eq:Nsp} and \ref{Eq:N2d}, the Fermi level of the metal can modulate the SPR signal. To modulate the Fermi level of a planar SPR film, we use a  phase sensitive SPR imaging system combined with an external signal generator which can apply a bias potential to the sensing surface according to the requirements as illustrated in Fig.~\ref{fig:1}(c). The sensing surface of a \qty{48}{nm} gold film is divided into two insulated cell: one as the working cell  connected with the signal generator and the other as the reference cell recording the light power fluctuation during the measurement. A region of interest (ROI) is selected in each cell to obtain the signal intensity of the region. In Fig.~\ref{fig:1}(c) , ROI 1 is in the working cell and ROI 2 in the reference cell. We use the differential of the intensities from both regions as the measured signal of which the noise from the light power fluctuation can be depressed. 

Fig.~\ref{fig:1}(d) shows an SPR signal under a linear bias potential scanning from \qtyrange{-0.5}{1}{\volt} at the rate of \qty{0.1}{\volt\per\second} in a \qty{30}{\milli\molar} potassium hydroxide (\ch{KOH}) electrolyte and the theoretical calculation of $n_{sp}$ under different Fermi level according to Eqs.~\ref{Eq:Nsp} and \ref{Eq:N2d}. The agreement, especially from \qtyrange{-0.5}{0.4}{\volt}, supports our previous prediction that we can use the SPR signal to indicate the surface electron density. The deviation of the signal from the theoretical calculation is because of the oxidation of the gold film when the bias potential exceeds \qty{0.5}{\volt} in solution. It should be noted that our previous conclusion \cite{liu2016potential}, supported by other studies \cite{foley2008surface,abayzeed2017sensitive}, that the SPR signal is proportional to the applied potential still holds because of $\ln\biggl( 1+\exp{\frac{E_f}{k_BT}}\biggr)\approx\frac{E_f}{k_BT}$ when $E_f >> k_BT$. For $T=298$ \unit{\K},  $k_BT$ is \qty{25.7}{\meV}.

Eq.~\ref{Eq:Nsp} also implies the ultra sensitivity of SPR to electron density variations at the sensing surface. SPR reduces the background electron density to $k_{sp}/k_F$, $10^{-3}$ approximately in our configuration. As an illustration, the background electron density is given by Eq.~\ref{Eq:N2d}, about \qty{4e5}{\per\square\um} at $E_f=100$ \unit{\meV} and the corresponding $n_{sp}$ is estimated by  \qty{400}{\per\square\um}. As a result, the signal from one electron increases from \qtyrange{e-6}{e-3}{\per\square\um} for SPR measurement.

Three technical factors block off the observation of this ultra electron density sensitivity: (1) the size of the detected spot, (2) the noise, and (3) the condition at which the individual molecules transfer their electrons to the sensing surface. In our prism-based SPR imaging system, the minimum detected spot is the area which a pixel is covering in the sensing surface, \qty{300}{\square\um} approximately ( \qty{25}{\um} $\times$ \qty{12}{\um} ). The number of the surface plasmon related electrons in the area, $\rho$, is about $10^5$ per pixel and the expected signal from one electron in the area, $1/\rho$, is $10^{-6}$ per pixel, which is still difficult to detect. A method is needed to further reduce these surface plasmon background electrons.

\begin{figure}[ht]
\centering
\includegraphics[width=0.9\textwidth]{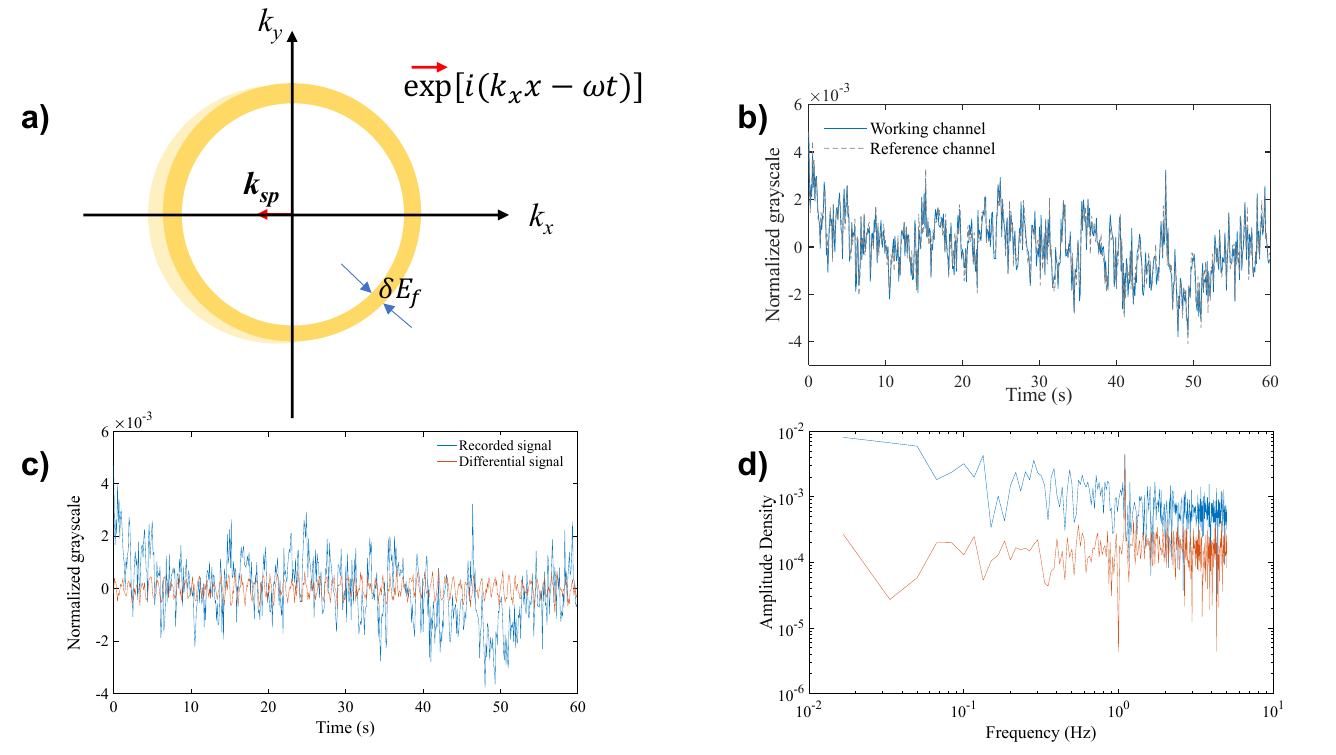}
\caption{(a)  The collective oscillation of surface plasmon electrons under an external bias potential, $\delta E_f$, mainly distribute in the ring induced from the Fermi level modulation in momentum space. (b)  The raw signals of ROI 1 and ROI 2. (c) The recorded signal of ROI 1 and the differential signal between ROI 1 and ROI 2 during 60 seconds in the time domain. (d) The amplitude spectrum density of the raw signal of ROI 1 and the differential signal in the frequency domain.}\label{fig:2}
\end{figure}

An investigation of  the surface plasmon electrons in momentum space  under an external bias potential paves the way to the reduction in Fig.~\ref{fig:2} (a). On the one hand, as we have pointed out, the surface plasmon occurs near the Fermi surface of the metal. On the other hand, an external potential applied to the metal can modulate the Fermi level of the metal. In momentum space, the Fermi surface under the potential modulation, $\delta E_f$, forms a ring with a certain number of electrons, $\delta n_s$. The estimation of $\delta n_s$  in vacuum can be obtained by the derivative of Eq.~\ref{Eq:N2d} with respect to $E_f$, or at a solid-liquid interface, it can be estimated by
\begin{equation} \label{Eq:dNs}
    \delta n_s = \frac{c\delta E_f}{e}
\end{equation}
where $c$ is the capacitance density of the interface. Under a surface plasmon condition, this ring also oscillates with $k_{sp}$.  Therefore, instead of measuring the total surface plasmon electrons, we focus on the electrons in the area of the ring, $\delta n_{sp}$. According to Eqs.~\ref{Eq:Nsp} and \ref{Eq:dNs}, we have:
\begin{equation} \label{Eq:dNsp}
    \delta n_{sp} = \frac{k_{sp}}{k_F}\frac{c\delta E_f}{e}
\end{equation}

For a gold electrode, the order of the capacitance density is \qty{10}{\uF\per\square\cm} \cite{moulton2004studies} 133and the corresponding electron density is about \qty{e4}{\per\square\um} approximately when $\delta E_f= $ \qty{100}{\meV}. The number of the surface plasmon electrons in response to $\delta E_f$ is estimated by \qty{10}{\per\square\um}. Therefore, under $\delta E_f$, the measured surface plasmon electrons in a pixel, $\rho$, is estimated by $10^3$ and one electron will induce $10^{-4}$ signal variation for $1/\rho$. 

To measure the signal variation under $\delta E_f$, we can apply a bias potential modulation function to the sensing surface. When excess electrons are introduced to the surface, the SPR signal in response to $\delta E_f$ will change. Especially, when the applied potential modulation function satisfies a trigonometric function, we can convert the SPR signal from the time-domain into the frequency-domain by Fourier transform of $\delta E_f$: $\delta\hat{n}_{sp}= \frac{k_{sp}}{k_F}\frac{c}{e}\mathcal{F}\{\delta E_f(t)\}$. 

In order to measure electron density variations on the order of $10^{-4}$ per pixel, noise is a second prevention,  among which the power fluctuation is dominant.  The large field of view of the imaging scheme can allow for the simultaneous recording of reflections from both the working cell and the reference cell. Because the reflections from both regions experience similar light power fluctuations during the measurement as shown in Fig.~\ref{fig:2}(b), the differential signal between the two regions can significantly reduce the effects of these fluctuations. To further improve the signal-to-noise ratio, sinusoidal potential modulations with an AC amplitude of \qty{200}{\mV} around different DC components at a frequency of \qty{1.1}{\Hz} are introduced to the working cell.

The comparison between the recorded signal in the working cell and the differential signal in Fig.~\ref{fig:2}(c) demonstrates a significant improvement in the signal-to-noise ratio for the latter. The power fluctuations in the recorded signal are measured to be $4\times 10^{-3}$, whereas the differential signal fluctuates less than $10^{-3}$. The modulation signal is barely noticeable in the raw signal but can be clearly distinguished in the differential signal. The corresponding amplitude density spectrum in Fig.~\ref{fig:2}(d) provides a comprehensive noise analysis. Prior to the depression scheme, the noise ranges from $10^{-3} $  $\sqrt{\unit{{\Hz}}}$  within the frequency band from \qtyrange{10}{1}{\Hz}, and approaches  $10^{-2}$ $\sqrt{\unit{{\Hz}}}$ within the low-frequency band from \qtyrange{1}{e-2}{\Hz}. After the differential, the noise is near the signal level, $10^{-4}$ $\sqrt{\unit{{\Hz}}}$, while the integration of the modulated signal at \qty{1.1}{\Hz} for one minute amplifies the measured signal to above $10^{-3}$ $\sqrt{\unit{{\Hz}}}$, providing the technique with an ability to analyze the electron density variation less than one electron. 

In the frequency-domain, we can estimate the SPR signal, $I_{AC}$, by
\begin{equation} \label{Eq:Iac}
     I_{AC} = KI_0\frac{\delta \rho}{\rho}\sqrt{\frac{\tau f}{2}}
\end{equation}
where $K$ is the constant determined by the optical setup, about $0.85$ in our configuration,  $I_0$ the background intensity of the surface, 157 grayscale levels during the experiment, $\tau$ integration period and $f$ is sampling frequency of CCD, optimized at \qty{10}{Hz}. 
\begin{figure}[ht]
\centering
\includegraphics[width=0.9\textwidth]{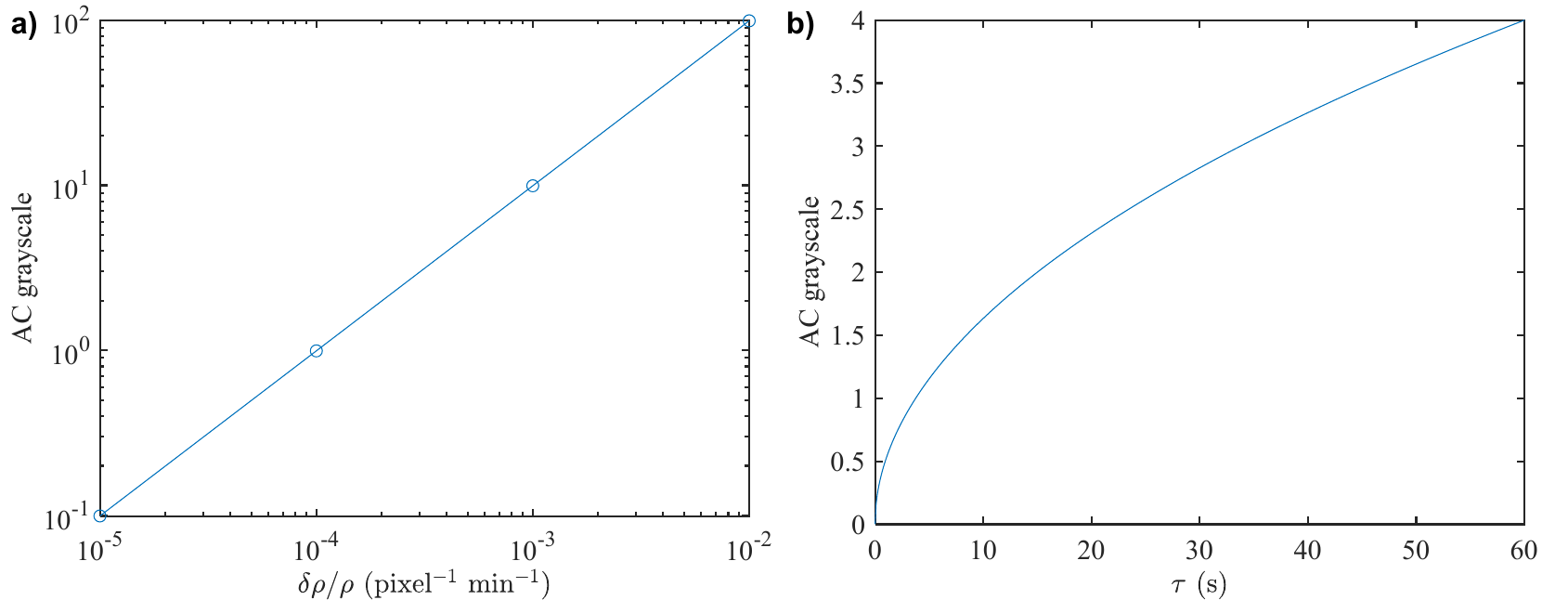}
\caption{(a) Frequency-domain signal caused by typical $\delta\rho/\rho$ per pixel during one minute; (b) Frequency-domain signal from one electron variation in a pixel vs. the duration time of the variation $\tau$. During the calculation, the refractive index of SF10 substrate is 1.723 and that of solution is 1.335. The electric constant of the gold is taken as -10.562+1.277i and the thickness of the gold film is 48 nm.}\label{fig:3}
\end{figure}

According to Eq.~\ref{Eq:Iac}, both electron density variations and corresponding integral time determine frequency-domain signals. For typical electron density variations from $10^{-5}$ per pixel per minute to $10^{-2}$ per pixel per minute, the frequency-domain signals change linearly from 0.1 grayscale levels to 100 grayscale levels in Fig.~\ref{fig:3}(a). For one electron variation in a pixel fitted in our configuration, $1/2500$ per pixel approximately, the frequency-domain signal during one-minute reaches 4 grayscale levels. In Fig.~\ref{fig:3}(b), the signal approaches to 4 grayscale levels by square root of the integral time, $\sqrt{\tau}$.

\begin{figure}[ht]
\centering
\includegraphics[width=0.9\textwidth]{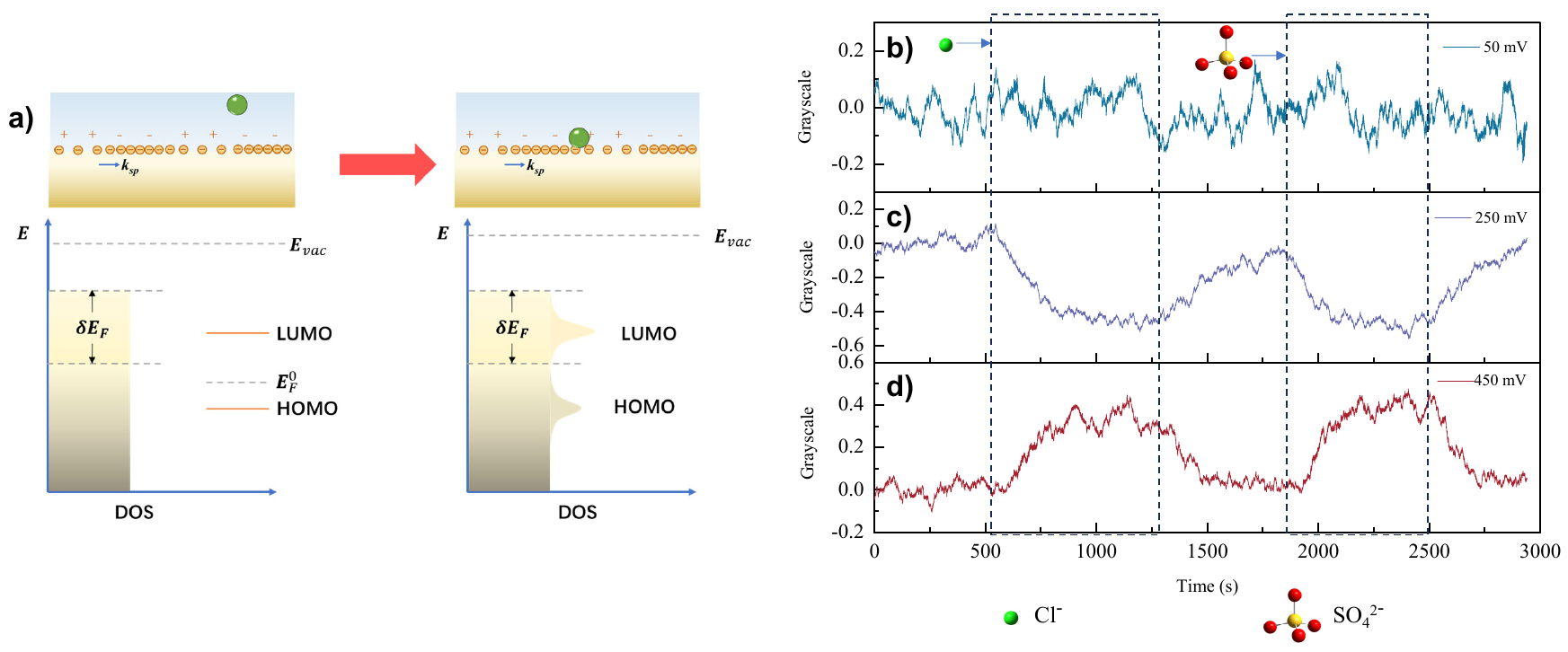}
\caption{(a) When the Fermi level of the metal is aligned with the degenerate molecular orbital level after the adsorption of an individual molecule, such as HOMO or LUMO, the DOS will change because of the electron transfer; (b), (c) and (d) The frequency-domain signal evolution of the adsorption and desorption of \qty{1}{\atto\molar} \ch{Cl^-} and \ch{SO4^2-} in \qty{30}{\milli\molar} \ch{KOH} at three different bias potential  modulations, \qty{50}{\mV} + \qty{200}{\mV}$\times\sin{(2\pi\times 1.1t)}$ in (b), \qty{250}{\mV} + \qty{200}{\mV}$\times\sin{(2\pi\times 1.1t)}$ in (c) and \qty{450}{\mV} + \qty{200}{\mV}$\times\sin{(2\pi\times 1.1t)}$ in (d). The dashed box represents the stages corresponding to the entry of the anions.}\label{fig:4}
\end{figure}
The condition at which the individual molecules transfer their electrons to the sensing surface is determined by the DC components of the potential modulations which set the Fermi level basis of the surface. The individual molecules transferring their electrons to the sensing surface occurs only when the frontier molecular orbital levels of the molecule, HOMO and LUMO, are aligned with the Fermi level of the surface as illustrated in Fig.~\ref{fig:4}(a) \cite{stadler2007fermi}. This energy level alignment changes the density of states (DOS) of the surface significantly around HOMO/ LUMO levels and subtly among other levels after the adsorption of an individual molecule. As a result, the electron density of the surface will be changed because of the alignment.

 To demonstrate the sensitivity of this ELA-SPRi imaging method, potassium chloride (\ch{KCl}) and potassium sulfate (\ch{K2SO4}) have been diluted by \qty{30}{\milli\molar} \ch{KOH} seperately to provide \qty{1}{\atto\molar} chloride anions (\ch{Cl^-}), and sulfate anions (\ch{SO_4^2-}). Both can be adsorbed on the gold surface \cite{tucceri1985surface}. In Fig.~\ref{fig:4}(b), (c) and (d), we have measured the adsorption and desorption of both anions at three different bias potential modulations. Although it is difficult to detect the processes under the modulation around \qty{50}{\mV}, we can distinguish the adsorption and desorption processes under the modulations around \qty{250}{\mV} and \qty{450}{\mV}.  For the Fermi level of the surface around \qty{50}{\mV}, the adsorption between the anions and the surface are physcial adsorption where molecules or atoms adhere to a surface through weak intermolecular forces, such as van der Waals forces or dipole-dipole interactions. However, when the Fermi level is aligned with the the frontier molecular orbital levels of the anions,  fractional electrons transfer between the molecule and the surface and the chemical adsorptions occur. For the alignment around \qty{250}{\mV} near the HOMO levels of both anions, the electrons transfer to the surface and the adsorption signals decrease while for the alignment around the LUMO levels,  \qty{450}{\mV}, the electrons transferring from the surface to the molecule increases the adsorption signals. It is interesting to notice that the opposite direction of the signal variations suggests the opposite electron transfer direction. 
 
It is also noticeable that traditionally we take chemical adsorptions irreversible. However, when the reactions occurs at single-molecule level, we observe the reversibility of the chemical adsorptions. When we wash the working cell with the electrolyte solution, \qty{30}{\milli\molar} \ch{KOH}, at \qty{1300}{\s} and \qty{2500}{\s}, the return of the signals in Fig.~\ref{fig:4}(c) and (d) indicates the desorption of the anions. Both anions introduce the similar intensity variations, about 0.4 grayscale levels, demonstrating about 0.1 e be introduced to the ROI on average. Thus, we can adapt the conventional Four-layer model to a three-layer model, glass substrate/ plasmonic metal film/ buffer medium, when the concentration of the detected molecules comes down to a single-molecule level. Instead of the molecular film thickness variation, the sparsely adsorbate molecules change the dielectric function of the plasmonic metal when the Fermi level is modulated around the frontier molecular orbital levels of the molecules. 

In conclusion, we have demonstrated the ultra electron density sensitivity for surface plasmons by investigating surface plasmon involved electrons from a solid-state perspective. The plasmonic nature can reduce the background electron density to the order of \qty{10}{\per\square\um} and be used to detect electron variations about 0.1 e. Based on this principle, we have developed an ELA-SPRi technique to measure the adsorption and desorption processes of individual anions at a planar gold electrode. This optical method, bypassing the molecular size restriction, provides a strategy to observing the electronic states of the individual molecules on a relative large structure and we plan to develop a single-molecule locating imaging technique based on the method. Furthermore, considering its compatibility with the conventional SPR-based technology and less limitations of the complicated sensing unit preparation, we believe ELA-SPRi will not only facilitate sensing in biophysical applications \cite{gooding2016single} but also be capable of analyzing individual chemical reactions in the field of catalysis engineering and energy \cite{wang2018heterogeneous,merryweather2021operando}.  
\begin{acknowledgments}
W. L. is grateful to Dr. Sixing Xu for his insightful suggestion that the detection response be from the electron density variation at the very beginning of this work and Miss Yu Miao for her constant urging the author to polish this work. This work was financed by National Key R\&D Program of China, Director Fund of Institute of Mechanics, and Fundamental Research Funds for the Central Universities.
\end{acknowledgments}

\bibliography{doc/latex/revtex/sample/aps/apssamp}

\end{document}